\documentclass{JHEP}
\input epsf
\usepackage{epsfig}
\usepackage{amssymb}
\usepackage[active]{srcltx}

\newcommand{\bea}{\begin{eqnarray}}
\newcommand{\eea}{\end{eqnarray}}
\newcommand{\bean}{\begin{eqnarray*}}
\newcommand{\eean}{\end{eqnarray*}}
\newcommand{\nn}{\nonumber \\}

\def\O #1{\overline{#1}}

\def\W #1{\widetilde{#1}}

\def\braket#1{\left\langle #1 \right\rangle}

\def\gb #1{ \left\langle #1 \right]}

\def\wt{\widetilde}

\def\a{{\alpha}}

\def\b{{\beta}}

\def\la{\lambda}

\def\vev{\braket}

\def\bvev#1{\left[ #1 \right]}
\def\Spaa{\vev}
\def\Spbb{\bvev}
\def\Spab{\gb}

\def\Label#1{\label{#1}%
  \smash{\hbox to0pt{\raise1ex\hbox{\tiny[#1]}\hss}}}

\preprint{
\\
{\tt hep-ph/}}
\title{Cross Section Evaluation by Spinor Integration II: The massive case in 4D}
\author{ Bo Feng$^{\diamond }$, Honghui Wang$^{\dagger}$
\footnote{Corresponding author.
 Email: wanggh06@yahoo.com.cn
}~~~~\\
$^\diamond$Center of Mathematical Science, Zhejiang University, Hangzhou 310027, China\\
$^{\dagger}$Zhejiang Institute of Modern Physics, Physics Department, Zhejiang University, Hangzhou 310027, China\\
}

\abstract{In this paper, we continue our study of  calculating the
cross section by the spinor method, i.e., performing the phase space
integration using the spinor method. We have focused on the case
where the physical momenta are massive and in pure 4D. We
established the framework of such a new method and presented several
examples, including two real progresses: $Z^0\to l^+ l^- H$ and $q\O
q\to f \O f H^0$. }

\begin{document}

\section{Introduction}
In the Tevatron collider and the LHC, multiple final states are
observed frequently. In order to check the standard model and
looking forward to finding new physics beyond the standard model
\cite{Albrow:2006rt,Bern:2008ef,Brooijmans:2008se, Morrissey:2009},
we need to explore the problem of how to calculate the cross section
efficiently and conveniently. In the past, the cross section is
evaluated in the 3-dimensional momentum space
\cite{Catani:1996,GehrmannDe Ridder:2003bm} and people have
developed quite mature numerical techniques. For the applications of
programs Madgraph, Pythia, AlpGen and Sherpa, the reader can check
references, for example,  \cite{Maltoni:2002}.

On the other hand, enormous progress have been made in the
evaluation of one-loop amplitudes \cite{Bern:2008ef}. One of such
progress is the unitarity cut method originally proposed in
\cite{Bern:1994zx,Bern:1994cg}. With the twistor program initiated
by Witten  \cite{Witten:2003nn},  the double cut phase space
integration has been reduced to algebraic manipulation through the
holomorphic anomily \cite{Cachazo:2004by,Britto:2004nj,
Britto:2005ha, Britto:2006sj}. Inspired by this simplification, in
our first paper \cite{Feng:2009ry}, we have explored how to apply
the spinor integration method to the evaluation of the cross section
for massless case. There are some obvious advantages comparing with
the momentum integration method. First, the three-dimension momentum
space integration can be reduced to just one-dimensional integration
and furthermore for the massless case, the integration region is
just $[0,1]$\footnote{The result of unitarity integration maybe
written as  one  Feynman parameter integration over rational
functions.}. Secondly, in the calculation, every step is manifestly
Lorentz invariant, thus we obtain compact analytic expressions.

Continuing our study for  the massless case, in this paper we focus
on the massive case. We will see that if all the mass is set to
zero, the massless case will be reproduced. Different from the
massless case, the integration variable $L$ is no longer a null
momentum. So we can't apply the spinor method directly. However this
problem have been solved in the unitarity cut method \cite{Anastasiou:2006jv,Anastasiou:2006gt}.
More accurately, we can write \bea \int d^4L=\int dz \int
d^4\ell\delta^+(\ell^2)(2\ell\cdot K);\quad L=\ell+zK, \eea where $K$ is a fixed
vector and $z$ is a real number. Through this decomposition, we
establish the general framework for massive case by  the spinor
method.

In our first paper, we have emphasized the advantages of using the
spinor method \cite{Feng:2009ry}. In the massive case, the
constrained three-dimensional momentum space integration still can
be reduced to an one-dimensional integration, plus possible Feynman
integrations. In every step, we get a scalar type of integrations,
which are Lorentz invariant. Furthermore, the integration region can
be written directly. Though it's not simply $[0,1]$ like the
massless case, it's only the simple functions of mass and energy.

The outline of this paper is as follows. In section 2,  we first
briefly review the 4D unitarity cut method. Then we take the
Faddeev-Popov trick to establish the general framework.

In section 3, we apply our method to the pure phase space
integration for two, three and four out-going particles as well as
some simple examples to demonstrate the main idea and feature. These
are the basis for practical and more complicated applications.

In section 4, we calculate two practical examples and summarize some
experience of performing the integrations.

A summary of our results with some comments is given in section 5.

\section{Framework to use spinor method}
In this section, we will setup the spinor integration method for
massive particles in 4D. Then we apply this method to the phase
space integration where the integration region (i.e, the $\int dx$)
of one dimensionless parameter is determined by the kinematical
discussion. This region corresponds to the boundary of the whole phase
space of outgoing momenta. One important difference, compared to the
massless case, is that the integration region will be functions of
masses of outgoing particles.

\subsection{The spinor integration method for massive cuts}
Here, we briefly review the spinor integration method for massive
cuts (or sometimes called the ``unitarity cut method")
\cite{Anastasiou:2006jv,Anastasiou:2006gt,Feng:2008gy}. The
Lorentz-invariant phase space (LIPS) of a massive double cut is
defined by inserting two $\delta$-functions representing the cut
conditions:
\bea \label{UPS}
 I=\int d^4~ \W\ell \delta^+(\W\ell^2-m_1^2)\delta((K-\W\ell)^2-m_2^2),
\eea
where $\W\ell$ is the internal loop momentum and $K$ the total
momentum through the unitarity cut. Because  $\W\ell$ is a massive
momentum, to use spinor integration method we need  decompose it as
\bea
\W\ell=\ell+zK,\quad \ell^2=0;\quad \int d^4\W\ell=
\int dzd^4\ell~\delta^+(\ell^2)(2\ell\cdot K).
\eea
where $\ell$ is a null 4-momentum, and can be expressed with spinor
variables as
\bea
\ell=tP_{\la\W\la},\quad P_{\la\W\la}=\la\W\la;\quad \int d^4\ell~\delta^+(\ell^2)=\int \Spaa{\la~\la}\Spbb{\W\la~\W\la}\int tdt.
\eea
Under this decomposition  Eq.(\ref{UPS}) becomes
\bea \label{UPS-1} I &=&\int dzd^4\ell\delta^+(\ell^2)(2\ell\cdot
K)\delta^+(z^2K^2+2zK\cdot \ell-m_1^2)\delta^+((1-2z)K^2-2K\cdot
\ell+m_1^2-m_2^2) \nn &=&\int
dz~((1-2z)K^2+m_1^2-m_2^2)\delta^+(z(1-z)K^2+z(m_1^2-m_2^2)-m_1^2)\nn
&&\int \Spaa{\la~\la}\Spbb{\W\la~\W\la}{(1-2z)K^2 +m_1^2-m_2^2\over
\Spab{\la|K|\W\la}^2},\quad t={(1-2z)K^2 +m_1^2-m_2^2\over
\Spab{\la|K|\W\la}}. \eea
In the first line of the result it depends only on the variable $z$, so we can
use the $\delta$-function to eliminate $z$ as follows
\bea z_\pm={(K^2+m_1^2-m_2^2)\pm \sqrt{\Delta[K,m_1,m_2]}\over
2K^2}~, \eea
where we have defined
\bea \Delta[K,m_1,m_2]
=(K^2-m_1^2-m_2^2)^2-4m_1^2 m_2^2~.\label{Delta} \eea
Between the two solutions of $z$, only one should be taken. To  see
that, we make a  kinematical analysis. Choose a center-of-mass
frame such that
\bean
K=(E>0,0,0,0),\quad \W\ell=(a,b,0,0),\quad K-\W\ell=(E-a,-b,0,0).
\eean
The mass-shell conditions require $a^2-b^2=m_1^2$ and $(E-a)^2-b^2=m_2^2$, so $a={(E^2+m_1^2-m_2^2)/2E}$.
In the decomposition $\W\ell=\ell+zK$, because the positive light cone with $\delta^+(\ell)$ have been chosen,  we can write
\bean
\ell=(|b|,b,0,0),\quad \W\ell=(|b|+zE,b,0,0)
\eean
Then $|b|+zE=a$. This means that only $z_-$ is retained\footnote{If $E<0$, we need $z_+$}.
In the following of this paper, we always refer to  $z$ as $z_-$ , if it's not explicitly illustrated.

Then Eq.(\ref{UPS-1}) becomes
\bea \label{UPS-2} I= \int
\Spaa{\la~\la}\Spbb{\W\la~\W\la}{(1-2z)K^2 +m_1^2-m_2^2\over
\Spab{\la|K|\W\la}^2},\quad t={(1-2z)K^2 +m_1^2-m_2^2\over
\Spab{\la|K|\W\la}}. \eea
Eq.(\ref{UPS-2}) is our final form for the spinor integration
with massive double cuts. For convenience we define
\bea z[K,m_1,m_2]={\alpha[K,m_1;m_2]-\beta[K;m_1,m_2]\over 2},\quad
t={\beta K^2\over \Spab{\la|K|\W\la}}, \eea
where
\bea \alpha[K,m_1;m_2]\equiv {K^2+m_1^2-m_2^2\over K^2},\quad
\beta[K;m_1,m_2]\equiv {\sqrt{\Delta[K,m_1,m_2]}\over K^2}.
~~~~~\label{def-ab}\eea
Notice that when $m_1=m_2=0$ we have
$\alpha=\beta=1$, thus reproducing the massless case. Finally the
original  $\W\ell$ can be parameterized as
\bea \label{W-ELL}
\W\ell=tP_{\la\W\la}+zK={K^2\over \Spab{\la|K|\W\la}}
\left[\beta\left(P_{\la\W\la}-{K\cdot P_{\la\W\la}\over K^2}K\right)+
\alpha{K\cdot P_{\la\W\la}\over K^2}K \right].
\eea
\subsection{Spinor integration method for the physical phase space integration}

Now, we explore how to apply the spinor integration method for massive cuts to the phase space integration.
Just like the massless case, when there are only two outgoing particles, spinor
integration method can be applied directly without any modification.
To see explicitly, just write the phase space of the cross section:
\bea I_2&=&\prod_{f=1,2}\int{d^4L_f\over
(2\pi)^3}\delta^+(L_f^2-m_i^2)(2\pi)^4\delta^4(K-\sum_{f=1,2}L_f)\nn
&\sim& \int d^4L_1\delta^+(L_1^2-m_1^2)\delta^+((K-L_1)^2-m_2^2),
\eea which is exactly the same (namely the show-up of two
$\delta$-functions)  as the spinor integration method given in
Eq.(\ref{UPS}).

Thing will be different when $n=3$, where the physical phase space
is given by:
\bea \label{I-3} I_3&=&\prod_{i=1}^3 \int {d^4 L_i\over (2\pi)^3}
\delta^+(L_i^2-m_i^2)(2\pi)^4\delta^4(K-\sum L_i)f(L_1,L_2, L_3)\nn
& = & \int {d^4 L_3\over (2\pi)^3} \delta^+(L_3^2-m_3^2)\int
{d^4L_2\over (2\pi)^2}\delta^+(L_2^2-m_2^2)\delta^+
((K-L_3-L_2)^2-m_1^2)f(L_2,L_3)\nn &=&\int {d^4 L_3\over (2\pi)^3}
\delta^+(L_3^2-m_3^2)\W f(L_3). \eea
The problem we meet here is just
the same as the massless case. The integration over $L_2$ with two
$\delta$-functions can be performed by the spinor integration method.
However, there is only one $\delta$-function in the integration over
$L_3$. In order to apply the spinor method recursively and
continuously, we need insert one more $\delta$-function like the
Faddeev-Popov method.

Similarly to the massless case \cite{Feng:2009ry}, we consider the
following integration
\bea I_x & \equiv & \int dx \delta( (x K- L_3)^2-m_3^2)  =  \int dx
\delta ( x^2 K^2- x(2K\cdot L_3)+L_3^2-m_3^2),\nonumber \eea
where the $\delta$-function has two solutions
\bea x_i & = & { (2K\cdot L_3)\pm \sqrt{(2K\cdot L_3)^2-4
K^2(L_3^2-m_3^2)}\over 2 K^2}~.\eea
Using the on-shell condition $L_3^2=m_3^2$, it reduces to
\bea  x_-=0,~~~~x_+= {2 K\cdot L_3\over K^2}~.\eea
We find that  $x_-=0$ is always a root. However, when $x=0$, we have
$\delta(L_3^2-m_3^2)$ which does not give an independent
$\delta$-function. So $x_-=0$ should be excluded from our
consideration. For another root $x_+$, from
$(K-L_3)^2=K^2+m_3^2-2K\cdot L_3\geq (m_1+m_2)^2$, we have
\bea 2K\cdot L_3\leq K^2+m_3^2-(m_1+m_2)^2,\eea
which gives the upper bound of $x_+$. For the lower bound,
considering the  center-of-mass frame where  $K=(E,0,0,0)$,
$L_3=(E_3,p,0,0)$ with $E_3^2-p^2=m_3^2$, we have $2K\cdot L_3\geq 2
E m_3$, i.e., $(2K\cdot L_3)^2\geq 4 K^2 m_3^2$.

Putting all consideration together we have
\bea I_x & \equiv & \int_{x_0}^{x_1} dx \delta( (x K- L_3)^2-m_3^2)
 = \int_{x_0}^{x_1} dx  {\delta(x-x_+)\over |2 x_+ K^2-(2K\cdot L_3)|}\nonumber \\
& = & {1\over |\sqrt{(2K\cdot L_3)^2-4
K^2(L_3^2-m_3^2)}|},~~~~\label{I-x-def}\eea
where
\bea  x_0[K,m_3]\equiv\sqrt{4 m_3^2\over
K^2},~~~~~x_1[K,m_3;m_t]\equiv {K^2+m_3^2-m_t^2\over
K^2}=\sqrt{x_0[K,m_3]^2+\Lambda[K;m_3, m_t]^2}~, \label{REGION}\eea
where $\Lambda[K;m_3, m_t]={\sqrt{\Delta[K,m_3,m_t]}/ K^2}$ with
$m_t=m_1+m_2$. Using $K^2\geq (m_1+m_2+m_3)^2$, it is easy to see
that $\Lambda[K;m_3, m_t]^2\geq 0$ and thus $x_1\geq x_0$.

Now Eq.(\ref{I-3}) can be written as
\bean
I_3&=& {1\over (2\pi)^3}\int d^4L_3 \delta^+(L_3^2-m_3^2){|\sqrt{(2K\cdot L_3)^2-4K^2 (L_3^2-m_3^2)}|}\int_{x_{0}}^{x_{1}}
 dx \delta((xK-L_3)^2-m_3^2)\W f(L_3).
\eean
Decomposing $L_3=\ell+zK$ with $\ell^2=0$, then
\bea
I_3&= &{1\over (2\pi)^3}\int dz d^4\ell\delta^+(\ell^2)(2\ell\cdot K)\delta^+(z^2K^2+2zK\cdot \ell-m_3^2){(2L_3\cdot K)}
\int_{x_{0}}^{x_{1}} dx \delta^+({x^2K^2-2xK\cdot L_3})\W f(\ell)\nn
&= &{1\over (2\pi)^3}\int_{x_{0}}^{x_{1}} dx{x K^2}\int dz(x-2z)K^2\delta^+(z(x-z)K^2-m_3^2)\nn
&&\hphantom{{1\over (2\pi)^3}\int_{x_{0}}^{x_{1}} dx}\times\int d^4\ell\delta^+(\ell^2) \delta^+(x(x-2 z)K^2-2 x K\cdot \ell)\W f(\ell).
\eea
One by-product of the above procedure is
\bea 2K\cdot L_3= x K^2~.~~\label{K-L3} \eea
By solving  the $\delta$-function $\delta^+(z(x-z)K^2-m_3^2)$
and the similar kinematical discussion as in Section 2.1, we get
\bea z & = &  { x K^2- \sqrt{ K^2(x^2 K^2-4 m_3^2)}\over
2K^2}={x\over 2}-{\sqrt{x^2-x_0^2}\over 2}, ~~~~ x-2z
=\sqrt{x^2-x_0^2}~. \eea
 Continue the evaluation as
\bea
 I_3 & = &{c\over (2\pi)^3} \int_{x_{0}}^{x_{1}} dx{x K^2} \int\Spaa{\la|d\la}\Spbb{\W\la|d\W\la}\int t dt \delta^+(x (x-2 z)K^2- xt
 \Spab{\la|K|\W\la})\W f(\la,\W\la,t)\nn
& = &{c\over (2\pi)^3} \int_{x_{0}}^{x_{1}} dx{x K^2} \int\Spaa{\la|d\la}\Spbb{\W\la|d\W\la} { (x-2z)K^2\over x\Spab{\la|K|\W\la}^2}\W f(\la,\W\la,t)
,~~~~t={ (x-2z)K^2\over\Spab{\la|K|\W\la}}\nn
& = & {c\over (2\pi)^3}\int_{x_{0}}^{x_{1}} dx
(K^2)^2\sqrt{x^2-x_0^2} \int\Spaa{\la|d\la}\Spbb{\W\la|d\W\la} {\W
f(\la,\W\la,t)\over\Spab{\la|K|\W\la}^2}, ~~~~t={
K^2\sqrt{x^2-x_0^2}\over\Spab{\la|K|\W\la}}, \eea
where $c=\pi/2$ is related to the Jacobi of changing integration variables and the way we have taken the residues.

Finally we arrive
\bea I_3&=&\int {d^4 L_3\over (2\pi)^3} \delta^+(L_3^2-m_3^2)\W
f(L_3)\nn &=&{\pi\over 2(2\pi)^3}\int_{x_{0}}^{x_{1}}
dx(K^2)^2\sqrt{x^2-x_0^2} \int\Spaa{\la|d\la}\Spbb{\W\la|d\W\la} {\W
f(\la,\W\la,t)\over\Spab{\la|K|\W\la}^2}, ~~~~t={
K^2\sqrt{x^2-x_0^2}\over\Spab{\la|K|\W\la}}~~~\Label{I3-form} \eea
This is our key setup in this paper. Notice that when
$m_1=m_2=m_3=0$, $x_0=0$, so  Eq. (\ref{I3-form}) reduces to
\bea I_3&=&\int {d^4 L_3\over (2\pi)^3} \delta^+(L_3^2)\W f(L_3)\nn
&=&{\pi\over 2(2\pi)^3}\int_{0}^{1} dx (K^2)^2 x
\int\Spaa{\la|d\la}\Spbb{\W\la|d\W\la} {\W
f(\la,\W\la,t)\over\Spab{\la|K|\W\la}^2},
~~~~t={K^2x\over\Spab{\la|K|\W\la}}. \eea
which is the familiar
massless case presented in \cite{Feng:2009ry}.

In the end, let us give a remark. The integration
region of $x\in [x_0, x_1]$ depends on the dynamical momentum $K$ as
well as mass parameters $m_3$ and $m_{total}$. Because this, the
roles of $m_3$ and $m_{total}$ are not obviously symmetric. If we
define
\bea x=\sqrt{x_0^2+\Lambda^2 u^2}~, \eea
the integration region of $u$ will be $u\in [0,1]$ which does not
depend on external momenta and masses anymore. Under this transformation we have
\bea I_3  & = & {\pi\over 2(2\pi)^3}\int_{0}^{1} du (K^2)^2 {\Lambda^3
u^2\over \sqrt{x_0^2+\Lambda^2 u^2}}
\int\Spaa{\la|d\la}\Spbb{\W\la|d\W\la} {\W
f(\la,\W\la,t)\over\Spab{\la|K|\W\la}^2}, ~~~~t={ K^2\Lambda
u\over\Spab{\la|K|\W\la}}.~~~\label{I3-form-1} \eea
and
\bea L_3 & = &{ K^2\Lambda u\over\Spab{\la|K|\W\la}}\la \W\la+ \left(
{\sqrt{x_0^2+\Lambda^2 u^2}\over 2}-{\Lambda u\over 2}\right) K. \eea
This transformation will become even simpler when $m_3=0$
where we get  a just linear transformation $x=\Lambda u$. Although Eq.
(\ref{I3-form-1}) may look simpler, however by some calculations we
find that Eq. (\ref{I3-form-1}) is, in general, not better than
Eq. (\ref{I3-form}) and readers can use anyone they like. In the later
part of this paper, we will use the form of Eq. (\ref{I3-form}).

\section{Simple examples}
In this section, we present some very simple examples to demonstrate
our method, especially  the integration region of $x$. We denote the
physical phase space integration of $n$ outgoing particles as
$I_n^{s\, \mathrm{or} \, m}(f;K)$, where $s$ stands for the spinor
method and $m$ the momentum method. The $K$ is the sum of momenta of
these $n$ particles and $f$ is a general function.
\subsection{The pure phase space integration with two outgoing particles}
This integral can be performed directly by the spinor method as we have analyzed in last section.

{\bf Spinor integration method}: The integration is given by
\bea
I_2^s(1;K) &=& \int {d^4L_2\over (2\pi)^3} {d^4L_1\over (2\pi)^3}\delta^+(L_2^2-m_2^2)
\delta^+(L_1^2-m_1^2)(2\pi)^4\delta^4(K-L_2-L_1)\\
&=&{1\over (2\pi)^2}\int d^4L_1\delta^+(L_1^2-m_1^2)\delta^+((K-L_1)^2-m_2^2).
\eea
According to Eq.(\ref{UPS-2}), one gets
\bea I_2^s(1;K)&=&{\pi\over 2(2\pi)^2}\int \Spaa{\la~\la}\Spbb
{\W\la~\W\la}{(1-2z)K^2 +m_1^2-m_2^2\over
\Spab{\la|K|\W\la}^2}={1\over (2\pi)^2}{\pi\over 2}\beta\nn & = &
{1\over 2(2\pi)^2}{\sqrt{(K^2-m_1^2-m_2^2)^2-4m_1^2 m_2^2}\over
K^2}, ~~~~ \label{I-2-S} \eea
 which is obviously symmetric between $m_1, m_2$.

{\bf Momentum integration method}: It is given by
\bea I_2^m(1;K) &=&
\int\frac{dL_1^3}{(2\pi)^32E_1}\frac{dL_2^3}{(2\pi)^32E_2}(2\pi)^4\delta^4(K-L_2-L_1).
\eea Taking the center-of-mass frame, where $K=(E,0,0,0)$ and
$L_1=(E_1,k_1,0,0)$, yields \bea I_2^m(1;K)&=&{1\over (2\pi)^2}\int
{k_1^2dk_1\over 2E_2}d\Omega \delta^+(E^2-2EE_1+m_1^2-m_2^2)
={1\over (2\pi)^2}{\pi\over 2}\beta. \eea
\subsection{The pure phase space integration with three outgoing particles}

{\bf Spinor integration method}: From Eq.(\ref{I-3}) with the result
$I_2^s(1;K)$ in previous subsection we have
\bea
I_3^s(1,K)&=&\int {d^4 L_3\over (2\pi)^3} \delta^+(L_3^2-m_3^2) I_2^s(1;K-L_3)\\
&=&{\pi\over 2(2\pi)^3}\int_{x_{0}}^{x_{1}} dx (K^2)^2
\sqrt{x^2-x_0^2} \int\Spaa{\la|d\la}\Spbb{\W\la|d\W\la} {
I_2^s(1;K-L_3)\over\Spab{\la|K|\W\la}^2}, \eea
where $I_2^s(1;K-L_3)$ depends on  $(K-L_3)^2$ only. But using
$L_3^2-m_3^2=0$ and $(xK-L_3)^2-m_3^2=0$, we can find
\bea \label{R-M-1} (K-L_3)^2&=& (1-x) K^2+m_3^2, \eea
which does not depend on $\la,\W\la$ at all.  Thus
\bea \label{I-3-s} I_3^s(1,K)&=&{\pi\over
2(2\pi)^3}\int_{x_{0}}^{x_{1}} dx K^2\sqrt{x^2-x_0^2}~
I_2^s(1;K-L_3), \eea
where $x_,x_1$ is given by (\ref{REGION}). This expression is
obviously symmetric between $m_1,m_2$, but not for $m_3$. However, it is easy
to check numerically that the final result is indeed symmetric among
all the masse parameters.

{\bf Momentum integration method}: The integration is
\bea
I_3^m(1;K) &= & \int {d^3L_1\over (2\pi)^32E_1} {d^3L_2\over (2\pi)^3 2E_2}{d^3L_3\over (2\pi)^3 2E_3}(2\pi)^4\delta^4(K-L_1-L_2-L_3)\\
&=&\int{d^3L_3\over (2\pi)^3 2E_3}I_2^m(1;K-L_3). \eea
In the center-of-mass frame, $K=(E,0,0,0)$, $L_3=(E_3,p,0,0)$ with
$E_3^2-p^2=m_3^2$, thus $(E-E_3)^2-p^2\geq(m_1+m_2)^2$, i.e.,
$E_3\leq(E^2+m_3^2-(m_1+m_2)^2/2E)$. Namely, the integration region
of $E_3$ is \bea m_3\leq E_3\leq {E^2+m_3^2-(m_1+m_2)^2\over 2E}~.
\eea Using this we have
\bea \label{I-3-M}
I_3^m(1;K) &= &{1\over (2\pi)^2}\int dE_3  \sqrt{E_3^2-m_3^2}I_2^m(1;K-L_3).
\eea
In order to show this is identical to Eq.(\ref{I-3-s}), we can make a transformation $2E_3/E \to x$. Then
\bea
m_3\leq E_3\leq {E^2+m_3^2-(m_1+m_2)^2\over 2E} &\to& x_0\leq x\leq x_1,\nn
2E\sqrt{E_3^2-m_3^2} &\to& K^2\sqrt{x^2-x_0^2}.
\eea
It's obvious to see that $I_3^m(1;K)=I_3^s(1,K)$.
\subsection{The pure phase space integration with four outgoing particles}
Here we will only present  the expression using spinor method. The
pure phase space is
\bea
I_4^s(1;K)&=&\int {d^4L_4\over (2\pi)^3} \delta^+(L_4^2-m_4^2)I_3^s(1;K-L_4).
\eea
Using the recursive method we get
\bea I_4^s(1;K)&=&{1\over
64(2\pi)^5}\int_{x^{(4)}_{0}}^{x^{(4)}_{1}} dx^{(4)}K^2
\sqrt{(x^{(4)})^2-(x^{(4)}_{0})^2}\nn & &
\int_{x^{(3)}_{0}}^{x^{(3)}_{1}}
dx^{(3)}(K-L_4)^2\sqrt{(x^{(3)})^2-(x^{(3)}_0)^2}~
I_2^s(1;K-L_3-L_4).~~~\label{phase4}\eea
where naively we have following boundary values
\bea  x_0^{(3)}= \sqrt{4 m_3^2\over (K-L_4)^2},~~~~~x_1^{(3)}=
{(K-L_4)^2+m_3^2-(m_1+m_2)^2\over (K-L_4)^2}.
\eea
However, similarly to Eq.(\ref{R-M-1}), we can find that
 \bea
(K-L_4)^2&=&(1-x^{(4)})K^2+m_4^2\nn
(K-L_4-L_3)^2&=&(1-x^{(3)})(K-L_4)^2+m_3^2=(1-x^{(3)})(1-x^{(4)})K^2+(1-x^{(3)})m_4^2+m_3^2
~~~\label{phase4-1}\eea
Thus we have
\bea x_0^{(4)}&=& \sqrt{4 m_4^2\over
K^2},~~~~~x_1^{(4)}={K^2+m_4^2-(m_1+m_2+m_3)^2\over K^2}\nn
 x_0^{(3)}&= &\sqrt{4 m_3^2\over (1-x^{(4)})K^2+m_4^2},~~~~~x_1^{(3)}={(1-x^{(4)})K^2+m_4^2+m_3^2-(m_1+m_2)^2\over (1-x^{(4)})K^2+m_4^2}
~~~\label{phase4-2}\eea
Putting (\ref{phase4-1}) and (\ref{phase4-2}) into (\ref{phase4}),
we get the analytic expression for the pure phase space of four
arbitrary massive particles
\bea I_4^{s}(1;K)
 &=&{1\over 64(2\pi)^5}\int_{x^{(4)}_{0}}^{x^{(4)}_{1}} dx^{(4)}\sqrt{K^2((x^{(4)})^2K^2-4m_4^2)}\int_{x^{(3)}_{0}}^{x^{(3)}_{1}} dx^{(3)}\nn
 &&\times {[((1-x^{(4)})K^2+m_4^2)((x^{(3)})^2((1-x^{(4)})K^2+m_4^2)-4m_3^2)]^{1\over 2}\over (1-x^{(3)})(1-x^{(4)})K^2+(1-x^{(3)})m_4^2+m_3^2}\nn
 &&\times [((1-x^{(3)})(1-x^{(4)})K^2+(1-x^{(3)})m_4^2+m_3^2-m_1^2-m_2^2)^2-4m_1^2m_2^2]^{1\over 2}.
 ~~~~\label{four-pure}
\eea
The expression is not obviously symmetric among $(m_1,m_2, m_3,
m_4)$ by our choice of the order of integrations. However, it is easy
to check by the numerical method that the final result is indeed
symmetric among all masses.

\subsection{The phase space integration of 3-outgoing particles with $f=(2L_1\cdot L_2)(2L_1\cdot L_3)$}
Here we calculate a relatively complicate example with $f=(2L_1\cdot L_2)(2L_1\cdot L_3)$.

{\bf Spinor integration method}: The integration can be directly written as
\bea I_3^s(f;K)&= & \int {d^4L_3\over (2\pi)^3}
\delta^+(L_3^2-m_3^2) I_2^s(f;K'),\quad K'=K-L_3, \eea
where using the momentum conservation, $f$ can be written as
\bea
f&=&(2L_1\cdot K'-2m_1^2)(2L_1\cdot (K-K')).
\eea
Using the Eq.(\ref{W-ELL}) with $K$ replaced by $K'$, we can
simplify $f$ further as
\bea f&=&(\alpha K'^2-2m_1^2)\left(\beta
K'^2{\Spab{\la|K|\W\la}\over \Spab{\la|K'|\W\la}}+(\alpha-
\beta)K'\cdot K-\alpha K'^2\right). \eea
 where
 \bea \alpha={K'^2+m_1^2-m_2^2\over
K'^2},\quad \beta={\sqrt{\Delta[K',m_1,m_2]}\over K'^2}.
~~~~\label{example32-a-b}\eea

Now  we calculate $I_2^s(f;K')$ using the simplified version $f$. It
is given by
\bea I_2^s(f;K') &=&{\pi\over 2(2\pi)^2}\alpha
\beta(K'^2-m_1^2-m_2^2)(K'\cdot K-K'^2). \eea
When we put it back  into $I_3^s(f;K)$, we need to
know that
 \bean
K'^2&=&(1-x) K^2+m_3^2\\
K'\cdot K&=&K^2-zK^2-{1\over 2}t\Spab{\la|K|\W\la}=K^2-{1\over
2}xK^2, \eean
where we have used the relation $ t={
(x-2z)K^2\over\Spab{\la|K|\W\la}} $. This means $I_2^s(f;K')$ does't
contain explicitly $\la$ and $\W\la$ for the spinor integration over $L_3$.
So we get immediately
\bea I_3^s(f;K)&=&{1\over
16(2\pi)^3}\int_{x_{0}}^{x_{1}} dx K^2\sqrt{x^2-x_0^2}\alpha
\beta((1-x) K^2+m_3^2-m_1^2-m_2^2)({1\over 2}xK^2-m_3^2). \eea
where $x_0,x_1$ and  $\a,\b$ are given by (\ref{REGION}) and (\ref
{example32-a-b}) respectively.

{\bf Momentum integration method}: The integration is
\bea
I_3^m(f;K)&= & \int {d^3L_3\over (2\pi)^32E_3} I_2^m(f;K').
\eea

$I_2^m(f;K')$ can be calculated as follows:
\bea
I_2^m(f;K')&=&{1\over (2\pi)^2}\int {d^3L_1\over 2E_1} \delta^+((K'-L_1)^2-m_2^2)(2L_1\cdot K'-2m_1^2)(2L_1\cdot (K-K')).
\eea
Choose a center-of-mass frame, such that $K'=(E',0,0,0)$, $L_1=(E_1,0,0,k_1)$ with $E_1^2-k_1^2=m_1^2$ and $K=(E,\vec{p})$.
The angle between  $\vec{p}$ and $\vec{L_1}$ is $\theta$. Then
\bea
I_2^m(f;K')&=&{1\over (2\pi)^2}\int {k_1^2dk_1\over 2E_1}\int _{-1}^{1}dy \int_0^{2\pi} d\varphi
\delta^+(E'^2-2E'E_1+m_1^2-m_2^2)\nn
&&\times(2E_1E'-2m_1^2)(2(EE_1-ypk_1)-2E_1E')\nn
&=&{1\over 2\pi}\int k_1dE_1
\delta^+(E'^2-2E'E_1+m_1^2-m_2^2)(2E_1E'-2m_1^2)(2EE_1-2E_1E')\nn
&=&{1\over 8\pi}{E'^2+m_1^2-m_2^2 \over
E'}{\sqrt{\Delta[E',m_1,m_2]}\over E'^2}(E'^2-m_1^2-m_2^2)(E-E').
\eea This is identical to $I_2^s(f;K')$ in the center-of-mass frame.
To  calculate $I_3^m(f;K)$ simply, we need to choose the
center-of-mass frame of $K$ which is not the one we have used for
$I_2^m(f;K')$, thus we need to write $I_2^m(f;K')$ as the
Lorentz-invariant form,  which is not so straightforward sometimes.

Here we use the Lorentz invariant form  $I_2^s(f;K')$ given by the
spinor method to go further. Taking the center-of-mass frame where
$K=(E,0,0,0)$, then \bea I_3^m(f;K)&=&{1\over 4(2\pi)^3}\int dE_3
\sqrt{E_3^2-m_3^2}{E^2-2EE_3+m_3^2+m_1^2-m_2^2 \over
(E^2-2EE_3+m_3^2)^2}{\sqrt{\Delta[E',m_1,m_2]}}\nn
&&\times(E^2-2EE_3+m_3^2-m_1^2-m_2^2)(EE_3-m_3^2). \eea It is equal
to $I_3^s(f;K)$,  which can be easily checked by making a transform
$2E_3/E \to x$ as in Section 3.2.

\section{Practical applications}

In our previous section we have done some simple examples. However,
these examples do not involve the real amplitudes. In this section we will
discuss the phase space integration of two simple real physics
progresses with three out-going particles. These two examples are
presented in the following two references:
 \cite{Kalyniak:1984} and \cite{N.Ng:1984}.
\subsection{$Z^0$ decays into lepton pairs and spin-0 bosons}

This example discuss the decay reaction
 \bea Z^0 \to l^+l^-H, \eea
where $l$ stands for the electron or muon with $m_l=0$ and $H$ for the
Higgs with  $m_H\neq0$. The invariant matrix element
squared is given by Eq.(2.10a) in \cite{Kalyniak:1984}. According to
the Glashow-Weinberg-Salam model (Eq.(3.1) and  Eq.(3.2) in
\cite{Kalyniak:1984}), the matrix element squared can be written as
\bea \overline{\sum_{pol}}|M|^2 &&={cM_Z^2\over
(Q^2-M_Z^2)^2}[Q^2M_Z^2+4(k\cdot l_+)(k\cdot l_-)],\quad c={2\over
3}(a^2+b^2)B_1^2, \eea
where $k$ is the total momentum of $Z^0$, $M_Z$ the mass of $Z^0$
and
\bean a={g\over \cos{\theta_W}}({1\over 4}-\sin^2\theta_W),\quad
b=-{g\over 4\cos{\theta_W}},\quad B_1={g\over
M_Z\cos{\theta_W}},\quad Q=l_++l_-. \eean
We evaluate this by first evaluating the phase space integration over
$l_+$ and $l_H$.
 From Eq.(\ref{I-2-S}), we can easily get
\bea I_2^s(\overline{\sum_{pol}};k-l_-)&=&{\pi\over 2(2\pi)^2}\int
\Spaa{\la \ d\la}\Spbb{\W\la \ d\W\la}{(k-l_-)^2-M_H^2 \over
\Spab{\la|k-l_-|\W\la}^2}\nn &&\times {cM_Z^2\over
(t\Spab{\la|l_-|\W\la}-M_Z^2)^2}(t\Spab{\la|l_-|\W\la}M_Z^2+2(k\cdot
l_-)t\Spab{\la|k|\W\la})\nn &=&{c\ \pi\over 2(2\pi)^2}\int \Spaa{\la
\ d\la}\Spbb{\W\la \ d\W\la} {((k-l_-)^2-M_H^2)^2\over
\Spab{\la|P_1|\W\la}^2\Spab{\la|P_2|\W\la}}\Spab{\la|R|\W\la}, \eea
where \bean
P_1&=&(k-l_-)-{(k-l_-)^2-M_H^2\over M_Z^2}l_-,\quad P_2=k-l_-\\
R&=&l_-+{2(k\cdot l_-)\over M_Z^2}k. \eean
Introducing one Feynman parameter we can continue to
 \bea
I_2^s(\overline{\sum_{pol}};k-l_-)&=&{c\ \pi\over
2(2\pi)^2}\int_0^1dy \int \Spaa{\la \ d\la}\Spbb{\W\la \
d\W\la}((k-l_-)^2-M_H^2)^2{2y\Spab{\la|R|\W\la}\over
\Spab{\la|P|\W\la}^3}\nn &=&{c\ \pi\over
2(2\pi)^2}\int_0^1dy((k-l_-)^2-M_H^2)^2{y2P\cdot R\over (P^2)^2},
~~~~\label{ex4.1-1}\eea
where
\bean
2P\cdot R&=&2k\cdot l_-+{2(k\cdot l_-)\over M_Z^2}(s y+u+k^2)\\
P^2&=&s y + u,\quad s={1\over M_Z^2}((2k\cdot
l_-)^2-(k^2-M_H^2)(2k\cdot l_-)),\quad u=k^2-2k\cdot l_-.\eean

Now we put (\ref{ex4.1-1}) into
 \bea
I_3^s(\overline{\sum_{pol}};k)&=&\int {d^4l_-\over (2\pi)^3}
\delta^+(l_-^2)I_2^s(\overline{\sum_{pol}};k-l_-). \eea
To continue, we exchange the integration order of $\int dy$ and
$\int d^4 l_-$. Using
 $2k\cdot l_-=t\Spab{\la|k|\W\la}=k^2x$ and performing the spinor integration,
 we finally arrive
  \bea I_3(\overline{\sum_{pol}};k)&=&{c\over
16(2\pi)^3}\int_0^{k^2-M_H^2\over k^2}dx (k^2x)^2 \int_0^1dy
(k^2-k^2x-M_H^2)^2 {y(1+{1\over M_Z^2}(s y+u+k^2))\over (sy+u)^2},
~~~~\label{4-1-1}\eea
where
\bean s&=&-{k^2x\over M_Z^2}(k^2(1-x)-M_H^2),\quad u=k^2(1-x).
\eean
%
We can perform the integral
over $y$ to yield
\bea I_3(\overline{\sum_{pol}};k)&=&{c\over
16(2\pi)^3}\int_0^{k^2-M_H^2\over k^2}dx
M_Z^2\left((k^2x+M_Z^2)\ln\left({1-{x(k^2(1-x)-M_H^2)\over
(1-x)M_Z^2}}\right)\right.\nn &&\left.-{(1-x)M_Z^2(k^2+M_Z^2)\over
(1-x)(k^2x-M_Z^2)-xM_H^2} +{k^2x(k^2(x-1)+M_H^2)\over
M_Z^2}-k^2-M_Z^2\right).~~~~\label{4-1-2} \eea
which can be integrated further to get analytic expression if one
wants.
Notice that $k^2=M_Z^2$, Eq. (\ref{4-1-2}) can be simplified further as
\bea
I_3(\overline{\sum_{pol}};k)&=&{2M_Z^4\over
384(\pi)^3}(a^2+b^2){g^2\over M_Z^2\cos^2{\theta_W}}\int_0^{1-{M_H^2\over M_Z^2}}dx
\left((x+1)\ln{{1+x^2+x({M_H^2\over M_Z^2}-2)\over
1-x}}\right.\nn
 &&\left.+{2(1-x)\over
1+x^2+x({M_H^2\over M_Z^2}-2)} +x^2+x({M_H^2\over
M_Z^2}-1)-2\right).~~~~\label{4-1-3}
\eea
Multiplied by the normalization factor $(2M_Z)^{-1}$ we have omitted in the calculation of
the cross section, the integrand of Eq. (\ref{4-1-3}) is just Eq. (3.4) in \cite{Kalyniak:1984}
by verifying $x=x_-$ and $\delta=M_H/M_Z$ in the center-of-mass frame.

\subsection{The production of Higgs bosons in $p\bar{p}$ collisions}

For the second real example, we consider the
quark-antiquark-annihilation mechanism $q\bar{q}\to f \bar{f}H^0$ in
\cite{N.Ng:1984}. The corresponding cross section and the matrix
element squared are respectively  given by Eq.(2.2) and  Eq.(2.3) in
\cite{N.Ng:1984}. We write the cross section as
\bea \label{HQ-3}
I_3^s(H^{\mu\nu}q^{\mu\nu};Q)&=&c\int d^4k \delta^+(k^2-m_f^2)\int
d^4\bar{k} \delta^+(\bar{k}^2-m_f^2)\int d^4h \delta^+(h^2-m_H^2)
\delta^4(Q-k-\bar{k}-h)H^{\mu\nu}q^{\mu\nu}\nn &=&c\int d^4k
\delta^+(k^2-m_f^2) I_2^s(H^{\mu\nu}q^{\mu\nu};Q'),\quad Q'=Q-k,
\eea
where $k(\bar{k})$ and $h$ are respectively the momentum of the
heavy quark(antiquark) $f(\bar{f})$ and the Higgs, $Q,q,\O q$ the
total momentum and momenta of particles $q, \O q$. We have absorbed
all the common constant factors including $\pi$-factor into $c$.
$H^{\mu\nu}q^{\mu\nu}$ is given by
\bea
H^{\mu\nu}q^{\mu\nu}&=&{32\over (2h\cdot \bar{k}+m_H^2)(2h\cdot
k+m_H^2)}\Bigg\{Q^2(Q\cdot h)^2\left[1+{(4m_f^2-m_H^2)Q^2\over
(2h\cdot \bar{k}+m_H^2)(2h\cdot k+m_H^2)}\right]\nn
&&\hphantom{{32\over (2h\cdot
\bar{k}+m_H^2)}}+\left[(Q^2+m_H^2-4m_f^2)+{2Q\cdot
h(4m_f^2-m_H^2)\over (2h\cdot \bar{k}+m_H^2)}\right]\left[{Q^2\over
2}m_f^2-2k\cdot q\ k\cdot \bar{q}\right]\nn &&\hphantom{{32\over
(2h\cdot \bar{k}+m_H^2)}}+\left[(Q^2+m_H^2-4m_f^2)+{2Q\cdot
h(4m_f^2-m_H^2)\over (2h\cdot k+m_H^2)}\right]\left[{Q^2\over
2}m_f^2-2\bar{k}\cdot q\ \bar{k}\cdot \bar{q}\right]\nn
&&\hphantom{{32\over (2h\cdot
\bar{k}+m_H^2)}}-(Q^2+m_H^2-4m_f^2)[2k\cdot q\ \bar{k}\cdot
\bar{q}+2k\cdot \bar{q}\ \bar{k}\cdot q-Q^2k\cdot \bar{k}]\Bigg\}.
\eea
Notice that \bean
2h\cdot \bar{k}+m_H^2&=&Q^2-2Q\cdot\bar{k}\\
2h\cdot k+m_H^2&=&Q^2-2Q\cdot k\\
2Q\cdot h&=&(Q^2-2Q\cdot k)+(Q^2-2Q\cdot \bar{k})\\
2k\cdot \bar{k}&=&(Q^2+m_H^2-2m_f^2)-(Q^2-2Q\cdot k)-(Q^2-2Q\cdot
\bar{k}). \eean
To simplify the calculation, we rearrange
$H^{\mu\nu}q^{\mu\nu}$ as
\bea \label{HQ}
H^{\mu\nu}q^{\mu\nu}&=&32\Bigg\{{1\over 4}Q^2\left[{Q^2-2Q\cdot
k\over Q^2-2Q\cdot \bar{k}}+2+{Q^2-2Q\cdot \bar{k}\over Q^2-2Q\cdot
k}\right]\nn &&+{1\over 4}(4m_f^2-m_H^2)(Q^2)^2\left[{1\over
(Q^2-2Q\cdot \bar{k})^2}+{2\over (Q^2-2Q\cdot \bar{k})(Q^2-2Q\cdot
k) }+{1\over (Q^2-2Q\cdot k)^2}\right]\nn &&+\left[{Q^2\over
(Q^2-2Q\cdot \bar{k})(Q^2-2Q\cdot k)}+{4m_f^2-m_H^2\over
(Q^2-2Q\cdot \bar{k})^2}\right]\left[Q^2m_f^2-2k\cdot q\ 2k\cdot
\bar{q}\right]\nn
&&-{Q^2+m_H^2-4m_f^2\over (Q^2-2Q\cdot \bar{k})(Q^2-2Q\cdot k)}[2k\cdot q\ 2\bar{k}\cdot \bar{q}-{1\over 2}Q^2((Q^2+m_H^2-2m_f^2)-(Q^2-2Q\cdot k))]\nn
&&-{(Q^2+m_H^2-4m_f^2)Q^2\over 2(Q^2-2Q\cdot k)}\Bigg\},
\eea
where we have used the symmetry between $k$ and $\bar{k}$.

Now we can start the calculation. First we will perform the phase space
integration over  $\bar{k}$ and $h$. Then we perform the left $k$ integration.

\subsubsection{The integration $I_2^s(H^{\mu\nu}q^{\mu\nu};Q')$}

First we simplify the input according to Eq.(\ref{W-ELL}), i.e.,
\bea \bar{k}&=&{ Q'^2\over \Spab{\la|Q'|\wt \la} }\left[\beta
(P_{\la\W \la}- {Q'\cdot P_{\la\W \la}\over Q'^2}Q')+\alpha {Q'\cdot
P_{\la\W \la}\over Q'^2}Q'\right],\nn
 Q^2-2Q\cdot \bar{k}&=&{
Q'^2\over \Spab{\la|Q'|\wt \la}}\Spab{\la|P_1|\W \la},\eea
with
 \bean P_1&=&-\beta(Q-{Q'\cdot Q\over Q'^2}Q')-(\alpha{Q'\cdot
Q\over Q'^2}-{Q^2\over Q'^2})Q'=-\beta Q_1+\alpha_1Q'. \eean
By checking   Eq.(\ref{HQ}), we find that there are four  nontrivial
integrations should be attacked. We do them one by one.

{\bf Type I}: $f_1=1/(Q^2-2Q\cdot \bar{k})$

The integration is
\bea
I_2^s(f_1;Q') &=  &\int \Spaa {\la~d\la }\Spbb {\wt \la~d\wt\la }{\beta \alpha_1\over \Spab{\la|\alpha_1Q'|\wt \la}\Spab{\la|P_1|\W \la}}\nn
&=&\int_0^1dy\int \Spaa {\la~d\la }\Spbb {\wt \la~d\wt\la }{\beta \alpha_1\over \Spab{\la|yP_1+(1-y)\alpha_1Q'|\W \la}^2}\nn
&=&\int_0^1dy{4\beta \alpha_1Q'^2\over -y^2\beta^2 \Sigma+4\alpha_1^2(Q'^2)^2},
\eea
where
\bean
\Sigma=(2Q\cdot Q')^2-4Q^2Q'^2.
\eean

{\bf Type II}: $f_2=Q^2-2Q\cdot \bar{k}$

The integral is \bea I_2^s(f_2;Q') &=&\int \Spaa {\la~d\la }\Spbb
{\wt \la~d\wt\la }{\beta Q'^2 \over \Spab{\la|Q'|\wt \la}^2} {
Q'^2\over \Spab{\la|Q'|\wt \la}}\Spab{\la|P_1|\W \la}\nn &=&\beta
Q'\cdot P_1=\beta \alpha_1 Q'^2. \eea

{\bf Type III}:  $f_3=1/(Q^2-2Q\cdot \bar{k})^2$

\bea  I_2^s(f_3;Q') &=  &\int \Spaa {\la~d\la }\Spbb {\wt \la~d\wt
\la }{\beta Q'^2 \over \Spab{\la|Q'|\wt \la}^2}{1\over (Q^2-2Q\cdot \bar{k})^2}\\
&=& \int \Spaa {\la~d\la }\Spbb {\wt \la~d\wt \la }{\beta /Q'^2\over \Spab{\la|P_1|\W \la}^2}\\
&=&{4\beta \over -\beta^2\Sigma+4\alpha_1^2(Q'^2)^2}.
\eea

{\bf Type IV}: $f_4=2\bar{k}\cdot \bar{q}/(Q^2-2Q\cdot \bar{k})$

Using
\bea
2\bar{k}\cdot\bar{q}
&=&{ Q'^2\over \Spab{\la|Q'|\wt \la} }\Spab{\la|P_2|\W \la},
\quad P_2=\beta (\bar{q}-{Q'\cdot \bar{q}\over Q'^2}Q')+\alpha {Q'\cdot  \bar{q}\over Q'^2}Q',
\eea
we get
\bea
 I_2^s(f_3;Q')
&=&\int \Spaa {\la~d\la }\Spbb {\wt \la~d\wt\la }{\beta Q'^2\over \Spab{\la|Q'|\wt \la}^2}{\Spab{\la|P_2|\W\la}\over \Spab{\la|P_1|\W\la}}\\
&=&\int_0^1dy\int \Spaa {\la~d\la }\Spbb {\wt \la~d\wt\la }{2\beta \alpha_1^2Q'^2(1-y)\Spab{\la|P_2|\W\la}\over \Spab{\la|yP_1+(1-y)\alpha_1Q'|\W\la}^3}\\
&=&\int_0^1dy{\beta \alpha_1^2Q'^2(1-y)[(-y\beta^2(1-{Q'\cdot Q\over Q'^2})+\alpha_1\alpha)2Q\cdot \bar{q}
-(y\beta^2{Q'\cdot Q\over Q'^2}+\alpha_1\alpha)2k\cdot \bar{q}]\over(-y^2\beta^2 {\Sigma\over 4Q'^2}+\alpha_1^2Q'^2)^2}
\eea

Now substituting  above four types of integrations into
$I_2^s(H^{\mu\nu}q^{\mu\nu};Q')$ and with some algebraic
manipulation, we can get
\bea {1\over
32}I_2^s(H^{\mu\nu}q^{\mu\nu};Q')&=&[{1\over 4}Q^2(Q^2-2Q\cdot
k)+{1\over 2}(Q^2)^2{6m_f^2-m_H^2\over Q^2-2Q\cdot k}-{Q^22k\cdot q\
2k\cdot \bar{q}\over Q^2-2Q\cdot k} -{1\over
2}Q^2(Q^2+m_H^2\nn
&&-4m_f^2)+{1\over
2}Q^2{(Q^2+m_H^2-4m_f^2)(Q^2+m_H^2-2m_f^2)\over Q^2-2Q\cdot
k}]\int_0^1dy{4\beta \alpha_1Q'^2\over -y^2\beta^2
\Sigma+4\alpha_1^2(Q'^2)^2}\nn &&-{(Q^2+m_H^2-4m_f^2)(2k\cdot q
)\over Q^2-2Q\cdot k}\int_0^1dy\beta \alpha_1^2Q'^2(1-y)\nn
&&{[(-y\beta^2(1-{Q'\cdot Q\over Q'^2})+\alpha_1\alpha)(2Q\cdot
\bar{q}) -(y\beta^2{Q'\cdot Q\over Q'^2}+\alpha_1\alpha)(2k\cdot
\bar{q})]\over(-y^2\beta^2 {\Sigma\over 4Q'^2}+\alpha_1^2Q'^2)^2}\nn
&&+{\beta (4m_f^2-m_H^2)\over
-\beta^2\Sigma+4\alpha_1^2(Q'^2)^2}[(Q^2)^2+4(Q^2m_f^2-2k\cdot q\
2k\cdot \bar{q})]+{(4m_f^2-m_H^2)\beta(Q^2)^2\over
4(Q^2-2Q\cdot k)^2}\nn
&&+{Q^2\beta
\alpha_1Q'^2-2Q^2\beta(Q^2+m_H^2-4m_f^2)\over 4(Q^2-2Q\cdot k)}+{1\over
2}\beta Q^2. \eea

\subsubsection{The integration $I_3^s(f;Q')$}

Now we do the left $k$-integration using (\ref{I3-form}) with
\bea x_0&=&\sqrt{4m_f^2\over Q^2},\quad
x_1={Q^2+m_f^2-(m_f+m_H)^2\over Q^2} \eea
and following relations
 \bean Q\cdot k&=&{x\over 2}Q^2,~~~~~
(Q-k)^2=(1-x)Q^2+m_f^2\nn
 2k\cdot q&=&{(x-2z_1)\Spab{\la|q|\W\la}Q^2\over  \Spab{\la|Q|\W\la}}+2z_1
Q\cdot q\nn
 2k\cdot\bar{q}&=&{(x-2z_1)\Spab{\la|\bar{q}|\W\la}Q^2\over
\Spab{\la|Q|\W\la}}+2z_1 Q\cdot \bar{q} \eean
From $I_2^s(H^{\mu\nu}q^{\mu\nu};Q')$, the terms containing $Q\cdot
k$ and $(Q-k)^2$ do not depend on $k$, thus can be done easily just
as in the example of the pure phase space integration with three outgoing particles.
 Then we
 need only perform the following two types of nontrivial integrations:
\bea I_3^s(2\bar{q}\cdot k;Q)&=&\int_{x_{0}}^{x_{1}} dx
(Q^2)^2\sqrt{x^2-x_0^2} \int \Spaa{\la|d\la}\Spbb{\W\la|d\W\la} {
1\over \Spab{\la|Q|\W\la}^2}({(x-2z_1)\Spab{\la|\bar{q}|\W\la}\over
\Spab{\la|Q|\W\la}}Q^2+2z_1 Q\cdot \bar{q})\nn
&=&\int_{x_{0}}^{x_{1}} dx
(Q^2)^2\sqrt{x^2-x_0^2}[(x-2z_1){Q\cdot \bar{q}\over Q^2}+{2z_1
Q\cdot \bar{q}\over Q^2}]\nn &=&\int_{x_{0}}^{x_{1}} dx
Q^2\sqrt{x^2-x_0^2}xQ\cdot \bar{q},\eea
\bea
I_3^s(2k\cdot q\ 2k\cdot \bar{q};Q)&=&\int_{x_{0}}^{x_{1}} dx (Q^2)^2\sqrt{x^2-x_0^2} \int
\Spaa{\la|d\la}\Spbb{\W\la|d\W\la} { 1\over \Spab{\la|Q|\W\la}^2}\nn
&&\times ({(x-2z_1)^2\Spab{\la|q|\W\la}\Spab{\la|\bar{q}|\W\la}\over  \Spab{\la|Q|\W\la}^2}(Q^2)^2
+2z_1 Q\cdot q{(x-2z_1)\Spab{\la|\bar{q}|\W\la}\over  \Spab{\la|Q|\W\la}}Q^2\nn
&&+2z_1 Q\cdot \bar{q}{(x-2z_1)\Spab{\la|q|\W\la}\over  \Spab{\la|Q|\W\la}}Q^2+2z_1 Q\cdot \bar{q}2z_1 Q\cdot q)\nn
&=&\int_{x_{0}}^{x_{1}} dx (Q^2)^2\sqrt{x^2-x_0^2} ((x-2z_1)^2{(2Q\cdot \bar{q})(2Q\cdot q)-Q^2(q\cdot \bar{q})\over 3Q^2}\nn
&&+2z_1 Q\cdot q(x-2z_1){Q\cdot \bar{q}\over Q^2}+2z_1 Q\cdot \bar{q}(x-2z_1){Q\cdot q \over Q^2}+{2z_1 Q\cdot \bar{q}2z_1 Q\cdot q\over Q^2})\nn
&=&\int_{x_{0}}^{x_{1}} dx Q^2\sqrt{x^2-x_0^2} \left({(2Q\cdot \bar{q})(2Q\cdot q)\over 3Q^2}(x^2Q^2-m_f^2)-{q\cdot \bar{q}\over 3}(x^2Q^2-4m_f^2)\right)
\eea

\subsubsection{The final result}
Substituting $I_2^s(H^{\mu\nu}q^{\mu\nu};Q')$, $I_3^s(2k\cdot q\ 2k\cdot \bar{q};Q)$ and $I_3^s(2\bar{q}\cdot k;Q)$ into Eq.(\ref{HQ-3}) yields
\bea
{1\over 32c}I_3^s(H^{\mu\nu}q^{\mu\nu};Q)
&=&\int_{x_{0}}^{x_{1}} dx \sqrt{x^2-x_0^2}\Bigg\{{1\over 2(1-x)}\int_0^1dy{\beta (2-2\alpha+\alpha x)\over -y^2\beta^2  (x^2-x_0^2)+(2-2\alpha+\alpha x)^2}T_1\nn
&&-{(Q^2+m_H^2-4m_f^2)\over Q^2(1-x)}\int_0^1dy{4\beta (1-y)(2-2\alpha+\alpha x)^2\over(-y^2\beta^2  (x^2-x_0^2)+(2-2\alpha+\alpha x)^2)^2}T_2\nn
&&+{\beta (4m_f^2-m_H^2)/Q^2\over -\beta^2 (x^2-x_0^2)+(2-2\alpha+\alpha x)^2}T_3+T_4\Bigg\},\label{4-2-1}
\eea
where
\bea
T_1&=&(Q^2x+m_H^2-4m_f^2)^2+(Q^2+m_H^2-4m_f^2)^2+4(Q^2+m_H^2-4m_f^2)m_f^2+2Q^2(6m_f^2-m_H^2)\nn
&&+{4\over 3}(q\cdot \bar{q}(x^2Q^2-4m_f^2)-(2Q\cdot
\bar{q})(2Q\cdot q)(x^2-{m_f^2\over Q^2})),\nn
T_2&=&(y\beta^2({x-{2m_f^2\over Q^2}})+\alpha (2-2\alpha+\alpha x))x(Q\cdot \bar{q})(Q\cdot q)\nn
&&+{1\over 6}(y\beta^2(2-x)+\alpha (2-2\alpha+\alpha x))(q\cdot \bar{q}(x^2Q^2-4m_f^2)-(2Q\cdot \bar{q})(2Q\cdot q)(x^2-{m_f^2\over Q^2})),\nn
T_3&=&(Q^2)^2+4Q^2m_f^2+{4\over 3}(q\cdot \bar{q}(x^2Q^2-4m_f^2)-(2Q\cdot \bar{q})(2Q\cdot q)(x^2-{m_f^2\over Q^2})),\nn
T_4&=&{\beta (4m_f^2-m_H^2)Q^2\over 4(1-x)^2}+{\beta(2-2\alpha+\alpha x)(Q^2)^2 -4(Q^2+m_H^2-4m_f^2)Q^2\beta\over 8(1-x)}+{1\over 2}\beta (Q^2)^2.
\eea
The corresponding parameters are
\bea
\alpha&=&{(1-x)Q^2+2m_f^2-m_H^2\over (1-x)Q^2+m_f^2},\quad \beta={[((1-x)Q^2-m_H^2)^2-4m_f^2m_H^2]^{1\over 2}\over (1-x)Q^2+m_f^2}.
\eea

\begin{figure}
\begin{center}
  \includegraphics[width=10cm]{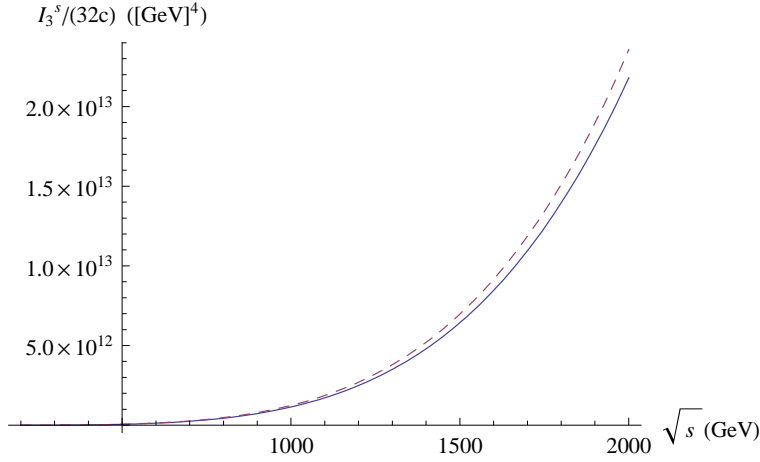}\\
  \caption{${1\over 32c}I_3^s$ as a function of $\sqrt{s}$ for the $p\bar{p}$ collision. The dashed and continuous curves respectively represent two sets of parameters: $m_H=10$ GeV, $m_f=4.5$ GeV and $m_H=30$ GeV, $m_f=35$ GeV.}\label{Example2-pic}
\end{center}
\end{figure}
In Fig. \ref{Example2-pic}, we display ${1\over 32c}I_3^s$ versus
the c.m. energy $\sqrt{s}$ of the $p\bar{p}$ by the numerical
method. Notice that the displayed ${1\over 32c}I_3^s$ is not the
real cross section since the dynamical factor $c$ given in the
original reference has not been included and the real cross section
is $I_3^s$ (so the decay behavior of $I_3$ at high energy  can not
be observed from this Figure).  Here, we emphasize two points.
First, by our spinor method, almost all calculations have been
reduced to reading out the residues of poles and making some
algebraic manipulations. Thus although the analytic expression looks
long, the calculation is kindly trivial.

Second, we can take appropriate integration order to simplify  the
process according to the structure of the integrand. Usually we
should  first perform the integrations over those variables, with
respect to which the structure of the integrand is relatively
simple. In this example, we have leave $k$ as the last integration
variable\footnote{Because the symmetry of $k$ and $\bar{k}$, it's
the same if we leave $\bar{k}$ as the last integration variable}.
 This is because the integrand Eq. (\ref{HQ}) does not contain $h$ explicitly.

 Notice that different integration ordering, i.e., integrating over $p_1$
 first and then $p_2$ or integrating over $p_2$ first and then $p_1$,
 will in general give  different-looking expressions. For
 example, in the expression of (\ref{four-pure}), we have fixed
 arbitrarily the ordering $m_1, m_2, m_3, m_4$. Different ordering
 will end up with different integration regions although the final result
 should be the same. Furthermore, if we have left one particle un-integrated
 while others have been integrated, then we
 will get the corresponding differential cross section for this
 particle. Thus different integration ordering will give different
 differential cross sections for different particles.

\section{Conclusion}
Originating from the application of the spinor method to the
massless case, in this paper, we have established the framework to
process the massive case. From the examples presented in the paper,
the advantages of our method is further manifested.

First, the manifestly Lorentz invariant form of the result in each step is gotten naturally.
This ensures that the recursive method can be applied conveniently especially when the number of outgoing particles is large.
In this process, we don't need take any specified reference frame as when using the momentum integration method.

Second, the integration regions can be written straightforward according to Eq.(\ref{REGION}), while with the momentum integration method, one has to pursue exhaustively
to specify those of many variables (for example, angles and module variables).
Note that in our method, though for the massive case the region  is not so simple as the massless case, it is only the functions of mass and energy.

Finally, the salient point is that the constrained three-dimensional momentum space integration is reduced to an one-dimensional
integration, plus possible Feynman integrations.
However, in this so large simplification, we just pay a little extra price, namely the integration over $\la$ and $\W\la$ which can be obtained
by reading out residues of corresponding poles.

In this paper, our new method have shown out the value of practical
calculations. As we have mentioned in the introduction, our method provides
compact analytic expressions for cross section. Thus we can
investigate the analytic structure using these expressions. We think
it is an interesting direction. Also, in this paper we have just
touched the tree-level result. It is our goal to combine these
analytic expressions with one-loop result to see if we can improve
current numerical NLO algorithm, especially the IR singularity
substraction. A regularization scheme is mandatory and we need consider the general D-dimensional case. All these questions will be our future projects.

{\bf Acknowledgement:}

We would like to thanks Dr. Kai Wang for providing us the references
for our two real examples and Prof. Luo for stimulating discussions.
The work is funded by Qiu-Shi funding as well as group funding
1A3000-172210115 from Zhejiang University, and Chinese NSF funding
under contract No.10875104.



\begin{thebibliography}{999}




\bibitem{Albrow:2006rt}
  M.~G.~Albrow {\it et al.}  [TeV4LHC QCD Working Group],
  arXiv:hep-ph/0610012.

\bibitem{Bern:2008ef}
  Z.~Bern {\it et al.}  [NLO Multileg Working Group],
  [arXiv:hep-ph/0803.0494].

\bibitem{Brooijmans:2008se}
  G.~H.~Brooijmans {\it et al.},
  arXiv:0802.3715 [hep-ph].




\bibitem{Morrissey:2009}
  David~E.~Morrissey, Tilman~Plehn, Tim~M.P.~Tait
  [arXiv:hep-ph/0912.3259]

\bibitem{Catani:1996}
  S.~Catani, M.H.~Seymour,
  Nucl.\ Phys.\ B {\bf 485}, 291 (1997)
  [Erratum-ibid.\ B {\bf 510} 503 (1998)]
  [arXiv:hep-ph/9605323]

\bibitem{GehrmannDe Ridder:2003bm}
  A.~Gehrmann-De Ridder, T.~Gehrmann and G.~Heinrich,
  Nucl.\ Phys.\  B {\bf 682}, 265 (2004)
  [arXiv:hep-ph/0311276].

\bibitem{Maltoni:2002}
  F.~Maltoni, T.~Stelzer,
  JHEP {\bf 0302}, 027 (2003)
  [arXiv:hep-ph/0208156];

  T.~Sjostrand, S.~Mrenna, P.~Skands,
  JHEP {\bf 0605}, 026 (2006)
  [axXiv:hep-ph/0603175];

    M.L.~Mangano, M.~Moretti, F.~Piccinini, R.~Pittau, A.D.~Polosa
  JHEP {\bf 0307}, 001 (2003)
  [arXiv:hep-ph/0206293];

  T.~Gleisberg {\it et al.}
  JHEP {\bf 0902}, 007 (2009)
  [arXiv:0811.4622];

  S.~Hoeche, F.~Krauss, S.~Schumann, F.~Siegert,
  JHEP {\bf 0905}, 053 (2009)
  [arXiv:0903.1219 [hep-ph]].




\bibitem{Bern:1994zx}
  Z.~Bern, L.~J.~Dixon, D.~C.~Dunbar and D.~A.~Kosower,
  Nucl.\ Phys.\ B {\bf 425}, 217 (1994)
  [arXiv:hep-ph/9403226].

\bibitem{Bern:1994cg}
  Z.~Bern, L.~J.~Dixon, D.~C.~Dunbar and D.~A.~Kosower,
  Nucl.\ Phys.\  B {\bf 435}, 59 (1995)
  [arXiv:hep-ph/9409265].

\bibitem{Witten:2003nn}
  E.~Witten,
  Commun.\ Math.\ Phys.\  {\bf 252}, 189 (2004)

\bibitem{Cachazo:2004by}
  F.~Cachazo, P.~Svrcek and E.~Witten,
  JHEP {\bf 0410}, 077 (2004)

\bibitem{Britto:2004nj}
  R.~Britto, F.~Cachazo and B.~Feng,
  Phys.\ Rev.\ D {\bf 71}, 025012 (2005)

\bibitem{Britto:2005ha}
  R.~Britto, E.~Buchbinder, F.~Cachazo and B.~Feng,
  Phys.\ Rev.\ D {\bf 72}, 065012 (2005)
  [arXiv:hep-ph/0503132].

\bibitem{Britto:2006sj}
  R.~Britto, B.~Feng and P.~Mastrolia,
  Phys.\ Rev.\ D {\bf 73}, 105004 (2006)

\bibitem{Feng:2009ry}
  B.~Feng, R.~Huang, Y.~Jia, M.~Luo, H.~Wang,
  Phys.\ Rev.\ D {\bf 81}, 016003 (2010)
  [arXiv:hep-ph/0905.2715].

\bibitem{Anastasiou:2006jv}
  C.~Anastasiou, R.~Britto, B.~Feng, Z.~Kunszt and P.~Mastrolia,
  Phys.\ Lett.\  B {\bf 645}, 213 (2007)
  [arXiv:hep-ph/0609191].


\bibitem{Anastasiou:2006gt}
  C.~Anastasiou, R.~Britto, B.~Feng, Z.~Kunszt and P.~Mastrolia,
  JHEP {\bf 0703}, 111 (2007)
  [arXiv:hep-ph/0612277].
\bibitem{Feng:2008gy}
   B.~Feng, G.~Yang
   Nucl.\ Phys.\ B {\bf 811} 305 (2009)
   [arXiv:hep-ph/0806.4016
   ]



\bibitem{Kalyniak:1984}
  P.~Kalyniak, John~N.~Ng, and P.~Zakarauskas,
  Phys.\ Rev.\ D {\bf 29}, 502(1984)
\bibitem{N.Ng:1984}
  John~N.~Ng, and Pierre~Zakarauskas,
  Phys.\ Rev.\ D {\bf 29}, 876(1984)
\end{thebibliography}
\end{document}